\documentclass[aip,jcp,reprint]{revtex4-2}
\usepackage[T1]{fontenc}
\usepackage[utf8]{inputenc}
\usepackage{amsmath,amssymb,bm,graphicx}
\usepackage{hyperref}
\allowdisplaybreaks
\begin{document}
\title{Static and dynamic density correlations of interacting active Brownian particles}

\author{Artur Soriani}
\thanks{These authors contributed equally to this work.}
\affiliation{Department of Physics and Materials Science, University of Luxembourg, L-1511 Luxembourg City, Luxembourg}

\author{Samuel Cameron}
\thanks{These authors contributed equally to this work.}
\affiliation{School of Mathematics and Statistics, The Open University, Walton Hall, Milton Keynes, MK7 6AA, United Kingdom}

\author{Tomer Markovich}
\affiliation{School of Mechanical Engineering, Tel Aviv University, Tel Aviv 69978, Israel}
\affiliation{Center for Physics and Chemistry of Living Systems, Tel Aviv University, Tel Aviv 69978, Israel}

\author{Elsen Tjhung}
\email{elsen.tjhung@open.ac.uk}
\affiliation{School of Mathematics and Statistics, The Open University, Walton Hall, Milton Keynes, MK7 6AA, United Kingdom}

\date{\today}

\begin{abstract}
We develop a stochastic Dean–Kawasaki field theory to predict the static structure factor and collective intermediate scattering function of interacting active Brownian particles.
Starting from the microscopic overdamped dynamics, we derive coupled stochastic equations for the density and polarization fields using an angular-moment closure.
Gaussian linearization then yields closed analytical expressions for static and dynamic density correlations in terms of propulsion speed, mean number density, interaction potential, translational and rotational diffusivities.
The coupling between density and polarization predicts a crossover from exponential relaxation to damped oscillations despite the absence of inertia. 
For Gaussian-core potential, the predictions contain no fitting parameters. 
Brownian dynamics simulations show quantitative agreement at low activity and long wavelengths, and reproduce the crossover behaviour. 
For Weeks--Chandler--Andersen potential, the nonintegrable core makes both the Fourier transform of the potential and the self-interaction correction term in the Dean--Kawasaki field theory ill-defined. 
We therefore introduce an approximation that replaces the interaction potential with a passive direct-correlation function fitted separately at each density and temperature. 
This approximation requires no additional activity-dependent fitting parameters.
\end{abstract}
\maketitle

%%%
\section{Introduction}\label{sec:introduction}
%%%

Active Brownian Particles (ABPs) provide a minimal model of self-propelled colloids: their positions evolve through persistent propulsion, interparticle forces, and translational diffusion, while rotational diffusion decorrelates their swimming directions. 
Even without attractive forces or aligning torques, repulsive ABPs can develop strong spatial correlations and undergo motility-induced phase separation (MIPS)~\cite{TailleurCates2008,Bialke2013,CatesTailleur2015,FodorMarchetti2018}. 
The resulting phase behaviour depends on the interaction potential, including the softness of the repulsion~\cite{LevisCodinaPagonabarraga2017,SanoriaChelakkotNandi2021}. 
Density correlations describe how particles are arranged in space and can be measured through the static structure factor in scattering experiments~\cite{deGennes1959}. 
At equilibrium, their long-wavelength limit is related to the compressibility. 
Their time dependence shows how density fluctuations relax at different length scales~\cite{hansen2006theory}. 
Static and dynamic density correlations therefore provide useful tests of coarse-grained theories that connect particle motion to collective behaviour.

Two central observables are the static structure factor $S(q)$ and the collective intermediate scattering function $F(q,t)$, which characterize the amplitude and temporal evolution of density fluctuations at wavevector $q$. 
In equilibrium, static correlations can be obtained from a Boltzmann distribution~\cite{hansen2006theory}.
For interacting ABPs, the stationary distribution is generally unknown. 
Moreover, persistent propulsion can produce an oscillatory relaxation despite completely overdamped microscopic dynamics. 
For a single ABP, these oscillations originate from directed motion before rotational diffusion randomizes its orientation~\cite{Kurzthaler2016}.
Indeed, this behaviour has been observed experimentally in dilute suspensions of catalytic Janus particles using particle tracking and differential dynamic microscopy~\cite{Kurzthaler2018}. 
These observations raise an important question: under what conditions do oscillations occur in an interacting suspension, and how do their onset, frequency, and damping depend on activity, density, and the interaction potential?

Before introducing our approach, we briefly review existing coarse-graining methods. 
Dynamical density functional theory (DDFT) connects microscopic particle dynamics to density and orientational fields~\cite{teVrugtLoewenWittkowski2020,SteffenoniFalascoKroy2017,BickmannWittkowski2020,teVrugtBickmannWittkowski2023,KafriTailleur2026}. 
This approach typically yields coupled deterministic equations for ensemble-averaged fields. 
The coefficients of the continuum theory are obtained from the radial and angular dependence of the pair distribution, fitted to simulation or experimental data~\cite{BickmannWittkowski2020,ZhangEtAl2021,BialkeLowenSpeck2013}.  
Collision-induced slowing down can then be represented by an effective density-dependent propulsion speed $v(\rho)$~\cite{Bialke2013,Speck2014,CatesTailleur2015,SolonCatesTailleur2015,CatesTailleur2013,SpeckEtAl2014}. 
This speed may be introduced phenomenologically or related to the anisotropic pair distribution~\cite{BialkeLowenSpeck2013}.

A related approach uses the many-body Fokker--Planck equation to approximate the many-body steady-state distribution at low activity.
Combined with passive two- and three-body correlations measured in simulations, this distribution yields predictions for active pressure and position--orientation correlations~\cite{CameronEtAl2023}. 
For active Ornstein--Uhlenbeck particles (AOUPs), the unified coloured noise approximation (UCNA) provides an approximate many-body steady-state distribution from which density profiles, correlations, and pressure can be calculated~\cite{MaggiEtAl2015,MarconiMaggi2015}.

Mode-coupling theory (MCT) uses the static structure factor, often obtained from experiments or simulations, together with microscopic dynamical parameters to predict density relaxation~\cite{Janssen2018}. 
It provides a self-consistent equation for the intermediate scattering function $F(q,t)$, with a memory kernel that accounts for how past density fluctuations affect their subsequent relaxation. 
For ABPs, this approach has been extended to include particle orientations~\cite{Liluashvili2017,Reichert2021,Szamel2019}.
Active MCT can also predict the nonequilibrium static structure factor $S(Q)$ from the passive structure factor $S_P(q)$ as the structural input~\cite{DebetsEtAl2024}.

The Dean--Kawasaki framework provides a microscopic route to fluctuating field theories and has been widely applied to active matter~\cite{Dean1996,Kawasaki1994,Illien_2025}. 
These applications use similar coarse-graining techniques but differ in their interaction models, approximations, and target observables. Linearized and mean-field treatments have predicted orientation-dependent pair structure in dilute ABPs~\cite{PoncetEtAl2021} and static pair correlations in active Ornstein--Uhlenbeck liquids, with the latter neglecting the polarization field~\cite{TociuEtAl2022}. 
Predictions for both $S(q)$ and $F(q,t)$ have been obtained for particles with quorum-sensing interactions, where the propulsion speed $v(\rho)$ depends on the sensed density, using a density-only description~\cite{DinelliEtAl2024}.
For ABPs with pairwise forces, related microscopic approaches have focused on MIPS coexistence through mechanical constitutive relations~\cite{OmarEtAl2023}, and on interfacial fluctuations~\cite{LangfordOmar2024}. 
Building on these developments, we retain stochastic density and polarization dynamics for pairwise-interacting ABPs to calculate stationary static and dynamic density correlations, including an explicit interaction-dependent criterion for the onset of oscillatory collective relaxation.

In this paper, we use the Dean--Kawasaki approach to derive the stochastic evolution of the position--orientation distribution of ABPs interacting through a pair potential. 
After closing the angular-moment hierarchy, we retain the coupled stochastic density and polarization fields and linearize about a homogeneous isotropic state.
This yields analytical expressions for the stationary $S(q)$ and $F(q,t)$, together with an explicit interaction-dependent criterion for the crossover from exponential relaxation to damped oscillations.
For Gaussian-core interactions, both observables follow directly from the microscopic parameters without fitting. 
For Weeks--Chandler--Andersen (WCA) interactions, we use a passive direct-correlation approximation.
The passive direct-correlation function is parameterised by four coefficients, fitted separately at each density and temperature but independent of activity.
The same passive four coefficients therefore determine both $S(q)$ and $F(q,t)$ without additional activity-dependent fitting parameters. 
We test both treatments against Brownian dynamics simulations.

%%%
\section{Methods}
%%%

At the microscopic level, we consider $N$ ABPs in a two-dimensional periodic square of side $L$. 
Particle $i$ has position $\mathbf{r}_i(t)$ and orientation $\hat{\boldsymbol{\nu}}_i(t)=(\cos\theta_i(t),\sin\theta_i(t))^T$, with $\theta_i(t)\in[0,2\pi)$. 
Its dimensionless overdamped Langevin dynamics are
\begin{align}
\dot{\mathbf r}_i &= v_0\hat{\boldsymbol\nu}_i - \sum_{j\ne i}\boldsymbol\nabla_i V(|\mathbf r_i-\mathbf r_j|) 
			+ \sqrt{2D_t}\boldsymbol\xi_i(t), \label{eq:rdot} \\
\dot\theta_i &= \sqrt{2D_r}\zeta_i(t). \label{eq:thetadot}
\end{align}
Here $v_0\geq0$ is the propulsion speed, $D_t>0$ and $D_r>0$ are the translational and rotational diffusivities, and $V(r)$ is an isotropic pair potential. 
$\boldsymbol{\nabla}_i$ is derivative with respect to $\mathbf{r}_i$.
$\boldsymbol\xi_i(t)$ and $\zeta_i(t)$ are independent Gaussian white noises with zero mean and delta correlations:
\begin{align}
\langle\xi_{i\alpha}(t)\xi_{j\beta}(t')\rangle &= \delta_{ij}\delta_{\alpha\beta}\delta(t-t'), \\
\langle\zeta_i(t)\zeta_j(t')\rangle &= \delta_{ij}\delta(t-t').
\end{align}
Their cross-correlation vanishes: $\langle\xi_{i\alpha}(t)\zeta_j(t')\rangle = 0$.
Greek indices $\alpha,\beta\in\{x,y\}$ denote Cartesian components, while Latin indices $i,j\in\{1,2,\dots,N\}$ label particles.
All quantities are dimensionless, with position, time, and potential measured in units of a reference length, time, and energy, respectively. 
The length and energy scales are set by the characteristic range and strength of the interaction potential $V(r)$.
Full details of the nondimensionalization are given in Appendix~\ref{sec:nondimensionalization}.

We parameterize the activity by $\mathrm{Pe}=v_0/D_r$, the persistence length measured in units of the reference length.
We also denote $\Pi=D_r/D_t$ as the ratio of the dimensionless rotational and translational diffusivities.
The dimensionless $D_t$ then measures the (effective) temperature relative to the interaction energy scale and controls the strength of noise.

%%%
\subsection{Fluctuating position-orientation distribution}\label{sec:field}
%%%

In this subsection, we derive the stochastic evolution for the position--orientation distribution $\psi(\mathbf{r},\theta,t)$ using the Dean--Kawasaki procedure~\cite{Dean1996,Kawasaki1994}.
We define the fluctuating distribution as
\begin{equation}
\psi(\mathbf r,\theta,t) = \sum_{i=1}^N \delta(\mathbf r-\mathbf r_i(t))\delta(\theta-\theta_i(t)). \label{eq:psi}
\end{equation}
Applying It\^o's lemma gives the distributional Dean--Kawasaki equation (see Appendix~\ref{sec:DK} for details)
\begin{widetext}
\begin{equation}
\partial_t\psi = 
-\boldsymbol\nabla\cdot
	\left[ v_0\psi\hat{\boldsymbol\nu} 
	- \psi \int d\mathbf r' \int d\theta' \, \psi(\mathbf r',\theta',t) \boldsymbol\nabla V(|\mathbf r-\mathbf r'|)
	+ \psi\mathbf{S}_0 + \sqrt{2D_t}\boldsymbol\pi \right]
	+ D_t\nabla^2\psi - \sqrt{2D_r}\partial_\theta\chi + D_r\partial_\theta^2\psi. \label{eq:psidot}
\end{equation}
\end{widetext}
$\boldsymbol{\nabla}$ is derivative with respect to $\mathbf{r}$ and $\partial_\theta$ is derivative with respect to $\theta$.
We have defined the particle-weighted noises as
\begin{align}
\boldsymbol\pi(\mathbf{r},\theta,t) &= \sum_{i=1}^N\boldsymbol\xi_i(t)\delta(\mathbf{r}-\mathbf{r}_i(t))\delta(\theta-\theta_i(t)) \\
\chi(\mathbf{r},\theta,t) &= \sum_{i=1}^N\zeta_i(t)\delta(\mathbf{r}-\mathbf{r}_i(t))\delta(\theta-\theta_i(t)).
\end{align}
For a fixed instantaneous particle configuration, $\boldsymbol{\pi}$ and $\chi$ have zero mean and covariance
\begin{align}
\langle\pi_\alpha(\mathbf r,\theta,t)\pi_\beta(\mathbf r',\theta',t')\rangle &= \psi(\mathbf{r},\theta,t)\delta_{\alpha\beta}\Delta_3, \\
\langle\chi(\mathbf r,\theta,t)\chi(\mathbf r',\theta',t')\rangle &= \psi(\mathbf{r},\theta,t)\Delta_3, \\
\langle\pi_\alpha(\mathbf r,\theta,t)\chi(\mathbf r',\theta',t')\rangle &= 0,
\end{align}
where $\Delta_3=\delta(\mathbf r-\mathbf r')\delta(\theta-\theta')\delta(t-t')$. 
The angular brackets $\left<\dots\right>$ denote averaging over noise/random kicks while holding the particles’ instantaneous positions and orientations fixed. 
The noise variance therefore depends on the local particle distribution $\psi(\mathbf r,\theta,t)$: more particles produce a larger variance.

The term $\mathbf S_0$ in Eq.~(\ref{eq:psidot}) corrects for the self-interaction introduced when the particle sum $\sum_i\sum_{j\ne i}$ is rewritten as an unrestricted density convolution, which appears in the second term on the right-hand side of Eq.~(\ref{eq:psidot}).
It therefore ensures that a particle does not interact with itself~\cite{Dean1996}. Explicitly,
\begin{align}
\mathbf{S}_0 = \boldsymbol{\nabla}_{\mathbf{s}}V(|\mathbf{s}|) \big|_{\mathbf{s}=\boldsymbol{0}} = \lim_{\mathbf{s}\rightarrow\boldsymbol{0}}V'(|\mathbf{s}|)\frac{\mathbf{s}}{|\mathbf{s}|}, \label{eq:self-interaction}
\end{align}
provided this limit exists. 
For a radial potential that is continuously differentiable near the origin and satisfies $\lim_{r\rightarrow0^+}V'(r)=0$, 
the correction term vanishes: $\mathbf S_0=\mathbf0$. This holds, for example, for the Gaussian-core potential $V(r)=e^{-r^2/2}$.
For a potential that is cusped or singular at the origin, this limit generally does not exist. 
One must then retain the explicit exclusion of self-pairs or regularize the potential near the origin.
A possible regularization replaces the singular core with a smooth core characterized by a small cutoff length~\cite{BeutlerEtAl1994}.

We define the number density and angular moment densities as
\begin{align}
\rho(\mathbf{r},t) &= \int_0^{2\pi}\psi(\mathbf{r},\theta,t) \, d\theta, \\
\rho(\mathbf{r},t) p_\alpha(\mathbf{r},t) &= \int_0^{2\pi}\hat\nu_\alpha(\theta)\psi(\mathbf{r},\theta,t) \, d\theta, \\
\rho(\mathbf{r},t) Q_{\alpha\beta}(\mathbf{r},t) &= \int_0^{2\pi}
	\left[ \hat\nu_\alpha(\theta)\hat\nu_\beta(\theta)-\frac{\delta_{\alpha\beta}}2 \right] \psi(\mathbf{r},\theta,t) \, d\theta,
\end{align}
where $\hat{\boldsymbol\nu}=(\cos\theta,\sin\theta)^T$.
In literature, $\rho\mathbf p$ is usually referred to as the polarization density, while $\mathbf p$ is referred to as the polarization or normalized polarization.
The normalized polarization $\mathbf p$ describes the local average particle orientation, while the symmetric traceless tensor $Q_{\alpha\beta}$ describes local nematic order.

Integrating Eq.~(\ref{eq:psidot}) over $\theta\in[0,2\pi)$ gives the time evolution for the number density $\partial_t\rho$, which couples to the polarization density $\rho\mathbf p$. 
Multiplying Eq.~(\ref{eq:psidot}) by $\hat\nu_\alpha(\theta)$ and integrating over $\theta\in[0,2\pi)$ gives the time evolution for the polarization density $\partial_t(\rho\mathbf p)$, which in turn couples to the nematic density $\rho Q_{\alpha\beta}$. 
Near the homogenous isotropic state, the nematic tensor scales as $Q_{\alpha\beta}\sim\mathcal{O}(\boldsymbol{\nabla}\mathbf{p})$~\cite{CatesTailleur2013,BialkeLowenSpeck2013}.
This will appear as a higher order gradient term in the time evolution for the polarization density $\partial_t(\rho\mathbf{p})$.
Thus, we may truncate $Q_{\alpha\beta}\simeq0$.
For $\mathbf S_0=\mathbf0$, the resulting coupled equations are
\begin{widetext}
\begin{align}
\partial_t\rho &= -\boldsymbol\nabla\cdot
	\left[ v_0\rho\mathbf p - \rho\int d\mathbf r'\,\rho(\mathbf r',t)\boldsymbol\nabla V(|\mathbf r-\mathbf r'|) 
		+ \sqrt{2D_t}\boldsymbol\Lambda \right] + D_t\nabla^2\rho, \label{eq:rhodot}\\
\partial_t(\rho p_\alpha) &= -\partial_\beta
	\left[ \frac{\rho v_0}{2}\delta_{\alpha\beta} - \rho p_\alpha\int d\mathbf r'\,\rho(\mathbf r',t)\partial_\beta V(|\mathbf r-\mathbf r'|)
		+ \sqrt{D_t}\Xi_{\alpha\beta} \right] 
	+ D_t \nabla^2(\rho p_\alpha) + \sqrt{D_r}\Upsilon_\alpha - D_r\rho p_\alpha, \label{eq:pdot}
\end{align}
\end{widetext}
where we have defined three new noise variables
\begin{align}
\Lambda_\alpha(\mathbf{r},t) &= \int_0^{2\pi}\pi_\alpha(\mathbf{r},\theta,t) \,d\theta,\\
\Xi_{\alpha\beta}(\mathbf{r},t) &= \sqrt2\int_0^{2\pi}\hat\nu_\alpha(\theta)\pi_\beta(\mathbf{r},\theta,t) \,d\theta,\\
\Upsilon_\alpha(\mathbf{r},t) &= \sqrt2\int_0^{2\pi}\left[\partial_\theta\hat\nu_\alpha(\theta)\right]\chi(\mathbf{r},\theta,t) \,d\theta.
\end{align}
The noises $(\Lambda_\alpha,\Xi_{\alpha\beta},\Upsilon_\alpha)$ have zero mean with covariance
\begin{align}
\left\langle \Lambda_{\alpha}(\mathbf{r},t)\Lambda_{\beta}(\mathbf{r}',t')\right\rangle
	&= \rho(\mathbf{r},t)\delta_{\alpha\beta}\Delta_2, \label{eq:Lambda-Lambda} \\
\left\langle \Xi_{\alpha\beta}(\mathbf{r},t)\Xi_{\gamma\delta}(\mathbf{r}',t')\right\rangle 
	&= \rho(\mathbf{r},t)\delta_{\alpha\gamma}\delta_{\beta\delta} \Delta_2, \\
\left\langle \Upsilon_{\alpha}(\mathbf{r},t)\Upsilon_{\beta}(\mathbf{r}',t')\right\rangle 
	&= \rho(\mathbf{r},t)\delta_{\alpha\beta}\Delta_2, \label{eq:Upsilon-Upsilon-correlation} \\
\left\langle \Lambda_{\alpha}(\mathbf{r},t)\Xi_{\gamma\delta}(\mathbf{r}',t')\right\rangle  
	&= \sqrt{2}\rho(\mathbf{r},t)p_\gamma(\mathbf{r},t)\delta_{\alpha\delta}\Delta_2, \label{eq:Lambda-Xi} \\
\left\langle \Xi_{\alpha\beta}(\mathbf{r},t)\Upsilon_{\gamma}(\mathbf{r}',t')\right\rangle &= 
	\left\langle \Lambda_{\alpha}(\mathbf{r},t)\Upsilon_{\beta}(\mathbf{r}',t')\right\rangle = 0, \label{eq:Xi-Upsilon}
\end{align}
where $\Delta_2=\delta(\mathbf{r}-\mathbf{r}')\delta(t-t')$.
These noises are multiplicative because their covariances depend on the local fields. 
The density and polarization current noises are also cross-correlated through $\rho\mathbf p$ in Eq.~(\ref{eq:Lambda-Xi}). 

%%%
\subsection{Gaussian theory}\label{sec:Gaussian}
%%%

Since the number density $\rho$ satisfies the continuity equation $\partial_{t}\rho+\boldsymbol{\nabla}\cdot\mathbf{J}=0$ [cf. Eq.~(\ref{eq:rhodot})], the average number density is constant:
\begin{equation}
\rho_{0}=\frac{1}{L^{2}}\int d\mathbf{r}\,\rho(\mathbf{r},t)=\frac{N}{L^{2}}.
\end{equation}
In Brownian dynamics simulations, the average density $\rho_0$ is fixed by the total number of particles $N$ and the system size $L\times L$. 

Let us consider a small perturbation around the homogenous isotropic state
\begin{equation}
\rho(\mathbf{r},t) = \rho_{0}+\delta\rho(\mathbf{r},t)
\quad\text{and}\quad 
\mathbf{p}(\mathbf{r},t) = \delta\mathbf{p}(\mathbf{r},t), \label{eq:rho-p-perturbation}
\end{equation}
where $\delta\rho/\rho_0$ and $|\delta\mathbf{p}|$ are much smaller than unity. 
We use the Fourier conventions
\begin{align}
\delta\rho(\mathbf r,t) &= \frac1L\sum_{\mathbf q}\delta\widetilde\rho_{\mathbf q}(t)e^{i\mathbf q\cdot\mathbf r}, \label{eq:deltarho-FT} \\
\delta\mathbf p(\mathbf r,t) &= \frac1L\sum_{\mathbf q}\delta\widetilde{\mathbf p}_{\mathbf q}(t)e^{i\mathbf q\cdot\mathbf r},\\
V(\mathbf{r}-\mathbf{r}') &= \frac1{L^2}\sum_{\mathbf q}\widetilde V_{\mathbf q}e^{i\mathbf q\cdot(\mathbf{r}-\mathbf{r}')},
\end{align}
with allowed wavectors $\mathbf{q}=\tfrac{2\pi}{L}(n_x,n_y)^T$, where $n_x,n_y\in\mathbb{Z}$.

Substituting the expansions (\ref{eq:rho-p-perturbation}) into Eqs.~(\ref{eq:rhodot}) and (\ref{eq:pdot}), 
we retain only terms belonging to the Gaussian theory. 
Specifically, we keep the linear deterministic terms and evaluate the field-dependent noise covariances at the homogeneous isotropic state, $\rho=\rho_0$ and $\mathbf p=\mathbf0$. 
The resulting noise is additive and Gaussian.
The discarded nonlinear and multiplicative-noise terms generate interaction vertices and may be included perturbatively to calculate higher-order corrections using Feynman diagrams.

Within this Gaussian theory approximation, density fluctuations couple only to the longitudinal polarization. 
Introducing the two-component vector
\begin{equation}
\mathbf{X}_\mathbf{q}(t) = 
\begin{pmatrix}
\delta\widetilde\rho_\mathbf{q}(t) \\
i\mathbf q\cdot\delta\widetilde{\mathbf p}_{\mathbf q}(t)
\end{pmatrix}, \label{eq:X}
\end{equation}
the Gaussian linearization of Eqs.~(\ref{eq:rhodot}) and (\ref{eq:pdot}) can be expressed as
\begin{equation}
\dot{\mathbf X}_{\mathbf q}(t) = -\mathbf M_{\mathbf q}\mathbf X_{\mathbf q}(t) + \mathbf Z_{\mathbf q}(t), \label{eq:Gaussian-theory}
\end{equation}
where 
\begin{equation}
\mathbf M_{\mathbf q} = 
\begin{pmatrix}
(D_t+\rho_0\widetilde V_{\mathbf q})q^2 & \rho_0v_0 \\
-v_0q^2/(2\rho_0) & D_r+D_tq^2
\end{pmatrix}. \label{eq:M} 
\end{equation}
$\mathbf{Z}_\mathbf{q}(t)$ in Eq.~(\ref{eq:Gaussian-theory}) is a Gaussian white noise with zero mean and covariance 
$\left< \mathbf{Z}_{\mathbf{q}}(t)\mathbf{Z}_{\mathbf{q}'}(t')^T \right> = \mathbf{L}_\mathbf{q}\delta_{\mathbf{q},-\mathbf{q}'}\delta(t-t')$,
where
\begin{equation}
\mathbf L_{\mathbf q} =
\begin{pmatrix}
2D_t\rho_0q^2 & 0 \\
0 & (D_tq^4+D_rq^2)/\rho_0
\end{pmatrix}, \label{eq:L}
\end{equation}
Each Fourier mode in Eq.~(\ref{eq:Gaussian-theory}) follows a two-component complex Ornstein--Uhlenbeck process, which can be solved exactly. 
The corresponding bare response propagator is $\mathbf{G}_0(\mathbf{q},\omega)=(\mathbf{M}_{\mathbf{q}}-i\omega\mathbf{I})^{-1}$.
Together with the Gaussian noise covariance $\mathbf{L}_\mathbf{q}$, this propagator provides the starting point for calculating perturbative corrections from the neglected nonlinear and multiplicative-noise terms using Feynman diagrams.

%%%
\subsection{Static and dynamic density correlations}\label{sec:correlations}
%%%

We define the time-dependent density correlation function as:
\begin{equation}
C(|\mathbf r-\mathbf r'|,t)=\langle\delta\rho(\mathbf r,\tau+t)\delta\rho(\mathbf r',\tau)\rangle_{\mathrm{ss}},
\end{equation}
where $\langle\cdots\rangle_{\mathrm{ss}}$ denotes time and ensemble averaging in the steady state. 
This function measures the correlation between density fluctuations at positions $\mathbf r$ and $\mathbf r'$, separated by a time interval $t$. 
Stationarity ensures that it depends on the time interval $t$ rather than the reference time $\tau$.

The static structure factor is the normalized spatial Fourier transform of the equal-time density correlation function:
\begin{equation}
S(q) = \frac{1}{\rho_0}\int d\mathbf{r} \, C(r,0) e^{-i\mathbf{q}\cdot\mathbf{r}} 
= \frac{1}{\rho_0}\left< \delta\widetilde{\rho}_{\mathbf{q}}\delta\widetilde{\rho}_{-\mathbf{q}} \right>_{\text{ss}}.
\end{equation}
Experimentally, $S(q)$ can be extracted from the scattering intensity of a monochromatic light source~\cite{ChaikinLubensky1995}.
The collective intermediate scattering function is the normalized spatial Fourier transform of the time-dependent density correlation function:
\begin{align}
F(q,t) &= \frac{1}{\rho_0}\int d\mathbf{r} \, C(r,t) e^{-i\mathbf{q}\cdot\mathbf{r}}  \\
&= \frac{1}{\rho_0}\left< \delta\widetilde{\rho}_{\mathbf{q}}(\tau+t))\delta\widetilde{\rho}_{-\mathbf{q}}(\tau) \right>_{\text{ss}}. \label{eq:Fq-def}
\end{align}
Thus, $F(q,0)=S(q)$. 
Experimentally, the normalized collective intermediate scattering function $F(q,t)/S(q)$ can be measured using dynamic light scattering or differential dynamic microscopy (DDM)~\cite{CerbinoTrappe2008}. 
DDM has also been used to measure oscillatory scattering functions in dilute Janus-particle suspensions~\cite{Kurzthaler2018}.

Let us now define the density-polarization covariance matrix as
\begin{equation}
\mathbf{C}_{\mathbf{q}}(t) = \left< \mathbf X_{\mathbf q}(\tau+t)\mathbf X_{\mathbf q}(\tau)^\dag \right>_{\text{ss}},
\end{equation}
where $\mathbf{X}_{\mathbf{q}}(t)$ is defined in Eq.~(\ref{eq:X}).
The intermediate scattering function is given by the element in the first row and first column of this matrix (and divided by $\rho_0$), i.e., $F(q,t)=C_{11,\mathbf q}(t)/\rho_0$.
Similarly, the static structure factor is given by $S(q)=C_{11,\mathbf{q}}(0)/\rho_0$.
Thus the task now is to find $\mathbf{C}_{\mathbf q}(t)$ and $\mathbf{C}_{\mathbf q}(0)$.

It can be shown that the equal-time covariance matrix $\mathbf{C}_{\mathbf{q}}(0)$ satisfies the following equation (see Appendix~\ref{sec:2d-OU} for details),
\begin{equation}
\mathbf M_{\mathbf q}\mathbf C_{\mathbf q}(0)+\mathbf C_{\mathbf q}(0)\mathbf M_{\mathbf q}^\dagger=\mathbf L_{\mathbf q}.
\end{equation}
Solving the resulting system of linear equations gives $\mathbf C_{\mathbf q}(0)$ and hence the static structure factor:
\begin{equation}
S(q) = \frac{1+\dfrac{\Pi^2\mathrm{Pe}^2}{2(\Pi+q^2)}\dfrac{\Pi+2q^2}{\Pi+2q^2+\rho_0D_t^{-1}\widetilde V_{\mathbf q}q^2}}
{1+\rho_0D_t^{-1}\widetilde V_{\mathbf q}+\dfrac{\Pi^2\mathrm{Pe}^2}{2(\Pi+q^2)}}, \label{eq:Sq}
\end{equation}
where $\Pi=D_r/D_t$ and $\text{Pe}=v_0/D_r$.
In the passive limit ($\text{Pe}=0$), Eq.~(\ref{eq:Sq}) reduces to the random-phase approximation~\cite{hansen2006theory}., $S(q)=[1+\rho_0\widetilde V_{\mathbf q}/D_t]^{-1}$
For passive Brownian particles, random-phase approximation has been shown to be accurate at high density $\rho_0$ or high temperature $D_t$~\cite{IkedaMiyazaki2011}.

The unequal time covariance matrix $\mathbf{C}_{\mathbf{q}}(t)$ has the formal solution (see Appendix~\ref{sec:2d-OU} for details):
\begin{equation}
\mathbf C_{\mathbf q}(t)=e^{-\mathbf M_{\mathbf q}t}\mathbf C_{\mathbf q}(0), \quad \text{for } t\ge0.
\end{equation}
Thus, the form of the intermediate scattering function $F(q,t)=C_{11,\mathbf q}(t)/\rho_0$ depends on whether the eigenvalues of $\mathbf M_{\mathbf q}$ are real and distinct, repeated, or complex conjugates. 
These eigenvalues are
\begin{align}
\lambda_\pm = {}& \frac{D_t}{2}\bigg[\Pi+2q^2+\rho_0D_t^{-1}\widetilde V_{\mathbf q}q^2 \nonumber\\
 & \pm\sqrt{(\Pi-\rho_0D_t^{-1}\widetilde V_{\mathbf q}q^2)^2-2\Pi^2\mathrm{Pe}^2q^2}\bigg]. \label{eq:lambda}
\end{align}
For distinct and real eigenvalues $\lambda_+\neq\lambda_-$, the intermediate scattering function is given by (see Appendix~\ref{sec:2d-OU})
\begin{align}
F(q,t) = \frac{D_tq^2-\lambda_-S(q)}{\lambda_+-\lambda_-}e^{-\lambda_+t} 
+\frac{\lambda_+S(q)-D_tq^2}{\lambda_+-\lambda_-}e^{-\lambda_-t}.\label{eq:Fq-1}
\end{align}
For a repeated real eigenvalue $\lambda$,
\begin{equation}
F(q,t)=e^{-\lambda t}\left[S(q)+(\lambda S(q)-D_tq^2)t\right]. \label{eq:Fq-2}
\end{equation}
For complex-conjugate eigenvalues $\lambda_\pm=\mu\pm i\Omega$, with $\Omega>0$,
\begin{align}
F(q,t)=e^{-\mu t}\biggl[ & S(q)\cos(\Omega t) \nonumber\\
 &+\frac{\mu S(q)-D_tq^2}{\Omega}\sin(\Omega t)\biggr]. \label{eq:Fq-3}
\end{align}
Thus, the intermediate scattering function can exhibit damped oscillations at a given nonzero wavevector $q$ when the eigenvalues $\lambda_\pm$ become complex conjugates. 
This occurs when the activity exceeds the critical value
\begin{equation}
\mathrm{Pe}>\mathrm{Pe}_c(q)=\frac{|\Pi-\rho_0D_t^{-1}\widetilde V_{\mathbf q}q^2|}{\sqrt2\Pi q}.\label{eq:Pe-critical}
\end{equation}
The microscopic dynamics remain entirely overdamped: the oscillations arise from persistent propulsion and density--polarization coupling. 
Oscillatory scattering functions have been predicted for isolated ABPs~\cite{Kurzthaler2016} and observed experimentally in dilute suspensions of catalytic Janus particles~\cite{Kurzthaler2018}. 
Oscillatory relaxation also occurs in transient density correlations within ABP-MCT~\cite{Liluashvili2017}.
Our stationary Gaussian theory provides an explicit interaction-dependent criterion for the onset of oscillatory collective density relaxation.

%%%
\section{Results and Discussions}\label{sec:simulations}
%%%

We use the BROWNIAN package in LAMMPS~\cite{THOMPSON2022108171} to simulate Eqs.~(\ref{eq:rdot}) and (\ref{eq:thetadot}). 
In all our simulations, we fix $\Pi=D_r/D_t=3$, which roughly corresponds to an isolated spherical particle with unit radius.
We then vary $D_t$, $\rho_0$, and $\mathrm{Pe}$ and measure the stationary $S(q)$ and $F(q,t)$. 
Numerical calculations of $S(q)$ and $F(q,t)$ from particles' positions $\{\mathbf{r}_i(t)\}$ are explained in Appendix~\ref{sec:Sq-Fq}.

\begin{figure}
\includegraphics[width=0.75\columnwidth]{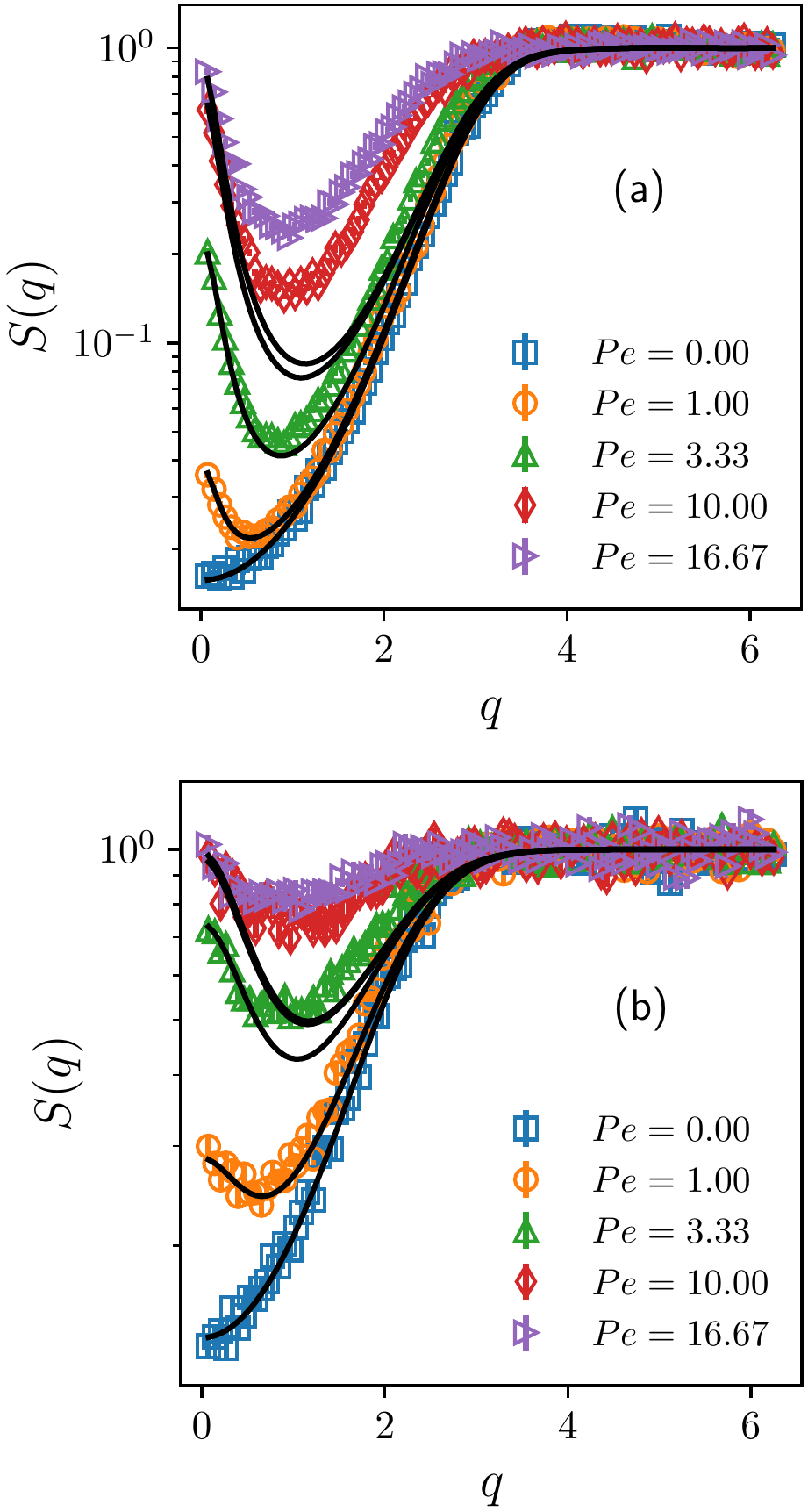}
\caption{Static structure factor for Gaussian-core ABPs at $\rho_0=1$: (a) $D_t=0.1$ and (b) $D_t=1$. Markers show simulations; black lines show Eq.~(\ref{eq:Sq}) without fitting parameters. In (b), the theoretical curves at $\mathrm{Pe}=10$ and $16.67$ nearly overlap.}
\label{fig:Gaussian-static}
\end{figure}

%%%
\subsection{Gaussian-core interactions}
%%%

The dimensionless Gaussian-core potential and its two-dimensional Fourier transform are
\begin{equation}
V(r)=e^{-r^2/2}, \qquad\widetilde V_{\mathbf q}=2\pi e^{-q^2/2}.
\end{equation}
For numerical efficiency, we truncate the interaction at $r_c=4$ in the simulations, while using the untruncated $\widetilde{V}_{\mathbf q}$ in our analytic Gaussian theory.
This truncation does not significantly affect the results reported here.
In this subsection, we fix the mean number density at $\rho_0=1$ and the translational diffusivity at $D_t=0.1$.
In our dimensionless units, $D_t$ measures the thermal noise strength relative to the characteristic interaction energy.
Choosing $D_t\ll1$ therefore suppresses thermally induced particle overlap.

Since $\widetilde V_{\mathbf q}>0$, both eigenvalues $\lambda_\pm$ have positive real parts for all $q>0$. 
In the case of complex eigenvalues $\lambda_\pm=\mu\pm i\Omega$, $\widetilde V_{\mathbf q}>0$ also implies that $\mu>0$ for all $q>0$.
Thus, the homogeneous isotropic state is linearly stable within the Gaussian theory approximation, and consequently the theory does not predict MIPS. 
Corrections to Gaussian theory can be investigated by restoring the nonlinear deterministic and multiplicative-noise terms from Eqs.~(\ref{eq:rhodot}) and (\ref{eq:pdot}), which generate interaction vertices in the Feynman diagrams.
Higher-order angular moments, such as $Q_{\alpha\beta}$, may also have to be included.

Fig.~\ref{fig:Gaussian-static} compares the simulated static structure factors with the Gaussian-theory predictions in Eq.~(\ref{eq:Sq}). 
Excellent quantitative agreement is observed at small wavevectors $q$ for all activities considered. 
This is consistent with the hydrodynamic limit underlying the angular-moment closure, i.e. we ignore the nematic order $Q_{\alpha\beta}$, which scales as the $\boldsymbol{\nabla}\mathbf p$.

\begin{figure}
\includegraphics[width=0.75\columnwidth]{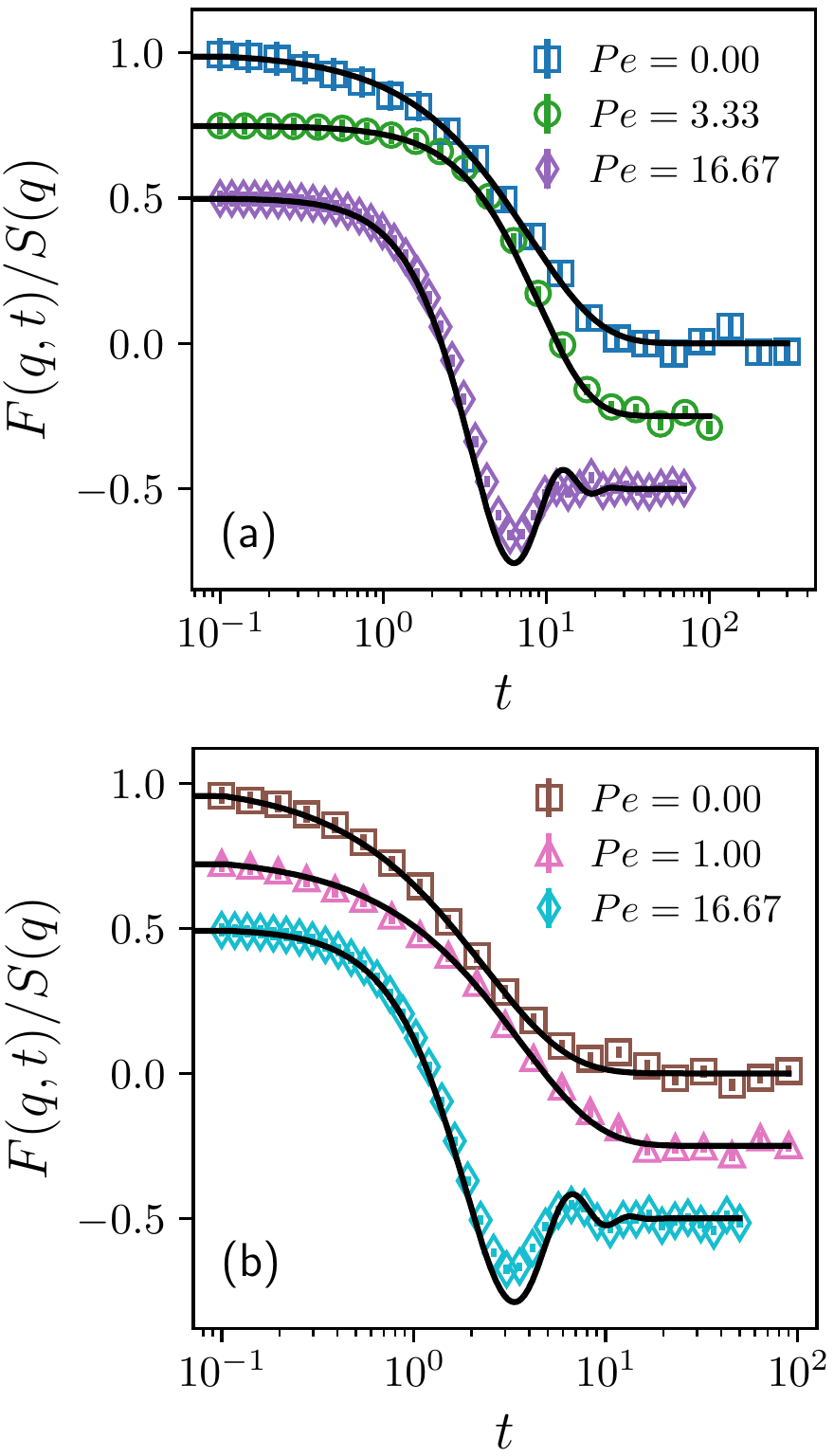}
\caption{Intermediate scattering function for Gaussian-core ABPs at $\rho_0=1$ and $D_t=0.1$: 
(a) $q\simeq0.14$ and (b) $q\simeq0.26$. 
Open markers show simulations. 
Black lines show analytic prediction from Eqs.~(\ref{eq:Fq-1})--(\ref{eq:Fq-3}) without fitting parameters. 
The thresholds for the onset of oscillations are $\text{Pe}_c(q=0.14)=3.0$ for (a) and $\text{Pe}_c(q=0.26)=1.0$ for (b).
By definition, $F(q,t=0)/S(q)=1$, however for clarity, some curves are shifted vertically.}
\label{fig:Gaussian-intermediate}
\end{figure}

Fig.~\ref{fig:Gaussian-intermediate} compares $F(q,t)$ divided by $S(q)$ at two different wavevectors: (a) $q=0.14$ and (b) $q=0.26$. 
The simulations reproduce the predicted crossover from exponential relaxation to damped oscillations as activity is increased.
The simulations also show close quantitative agreement with the theory given in Eqs.~(\ref{eq:Fq-1})--(\ref{eq:Fq-3}) using (\ref{eq:Sq}). 
Eq.~(\ref{eq:Pe-critical}) gives the thresholds $\mathrm{Pe}_c\simeq3.0$ at $q\simeq0.14$ and $\mathrm{Pe}_c\simeq1.0$ at $q\simeq0.26$. 
Just above these thresholds, the oscillations are strongly damped and difficult to distinguish from statistical noise in the simulations. 
We therefore use a higher activity, $\mathrm{Pe}=16.67$, in Fig.~\ref{fig:Gaussian-intermediate} to make the oscillations clearly visible.
Both $S(q)$ and $F(q,t)$ are determined by the same microscopic parameters, so the analytical predictions for the Gaussian-core potential contain no fitting parameters.

\begin{figure*}
\includegraphics[width=\textwidth]{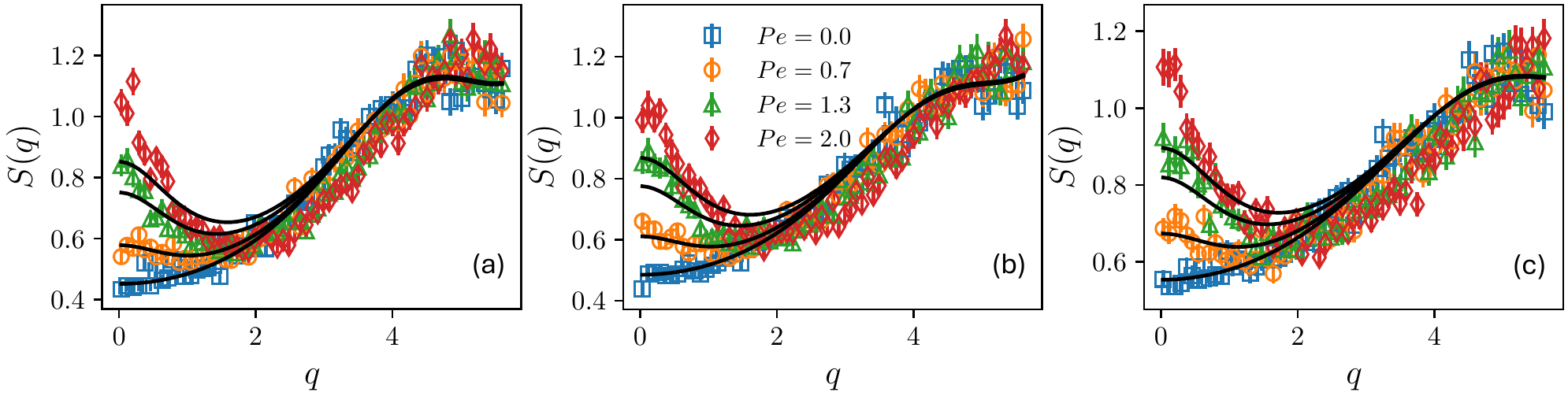}
\caption{Static structure factors for WCA ABPs at $\rho_0=0.2$, with (a) $D_t=0.05$, (b) $D_t=0.5$, and (c) $D_t=5$. 
Solid black lines show the analytical predictions from Eq.~(\ref{eq:Sq}) and (\ref{eq:passive-replacement}), with the coefficients in Eq.~(\ref{eq:cP}) fitted to passive simulations ($\mathrm{Pe}=0$) and held fixed as activity is varied. 
No additional fitting parameters are introduced.}
\label{fig:WCA-static}
\end{figure*}

\begin{figure*}
\includegraphics[width=\textwidth]{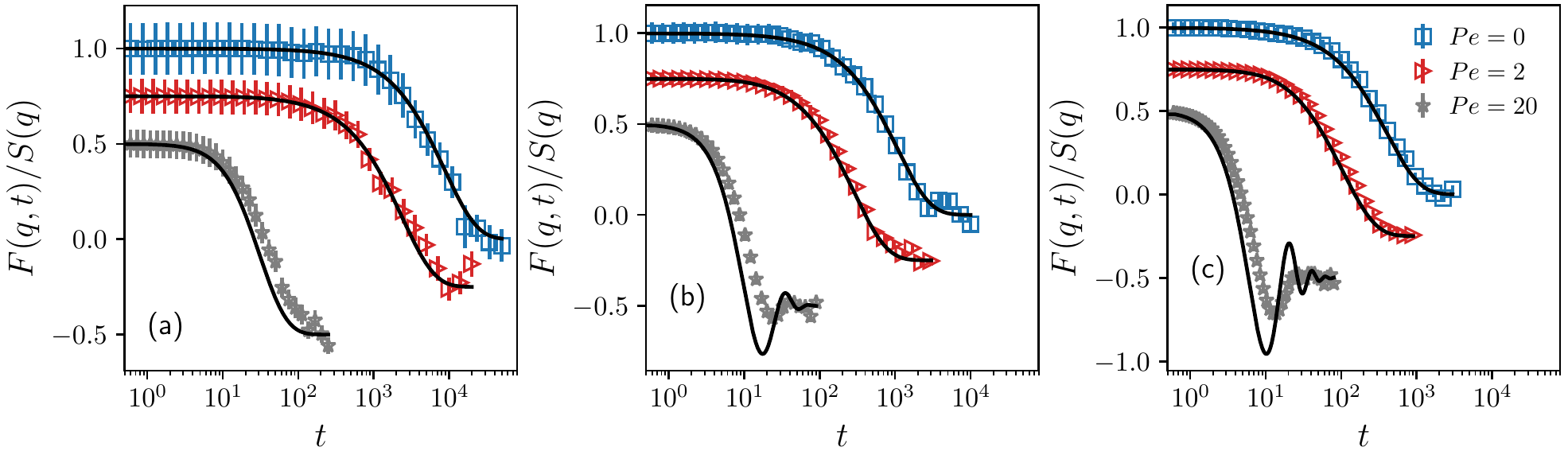}
\caption{Intermediate scattering function for WCA ABPs at $\rho_0=0.2$ and $D_t=0.05$: (a) $q=0.0320$, (b) $q=0.0915$, and (c) $q=0.149$.
Solid black lines show the analytical predictions from Eqs.~(\ref{eq:lambda})--(\ref{eq:Fq-3}) and (\ref{eq:passive-replacement}), with the coefficients in Eq.~(\ref{eq:cP}) fitted to passive simulations ($\mathrm{Pe}=0$).
The critical $\text{Pe}_c$ for the onset for oscillations are $\text{Pe}_c(q=0.0320)=22$, $\text{Pe}_c(q=0.0915)=7.7$, and $\text{Pe}_c(q=0.149)=4.7$.}
\label{fig:WCA-intermediate}
\end{figure*}

%%%
\subsection{WCA interactions and passive structural input}
%%%

The WCA potential~\cite{WeeksChandlerAndersen1971} is
\begin{equation}
V(r)=\begin{cases}4r^{-12}-4r^{-6}+1,&r<2^{1/6},\\0,&r\geq2^{1/6}.\end{cases}
\end{equation}
Its singular core means that the Fourier transform $\widetilde{V}_{\mathbf{q}}$ does not exist.
Furthermore, the self-interaction correction term $\mathbf{S}_0$ in Eq.~(\ref{eq:psidot}) is also ill-defined. 
Together, these prevent direct application of Eq.~(\ref{eq:Sq}) and (\ref{eq:lambda}).

We therefore introduce an approximation which replace $\widetilde{V}_{\mathbf{q}}$ in Eq.~(\ref{eq:Sq}) and (\ref{eq:lambda}) by a quantity, called the direct correlation function $\widetilde{c}_P(q)$ (in Fourier space).
More precisely, we replace
\begin{equation}
\frac{\widetilde V_{\mathbf q}}{D_t} \rightarrow -\widetilde c_P(q) \label{eq:passive-replacement}
\end{equation}
in Eqs.~(\ref{eq:Sq}) and (\ref{eq:lambda}).

The direct correlation function $\widetilde{c}_P(q)$ is obtained from the passive structure factor $S_P(q)$ at a given density $\rho_0$ and diffusivity $D_t$ through the Ornstein--Zernike relation~\cite{hansen2006theory}:
\begin{equation}
S_P(q) = \frac1{1-\rho_0\widetilde c_P(q)}, \qquad q>0.\label{eq:SP}
\end{equation}
Since the system is isotropic, we expect $\widetilde{c}_P(q)$ to contain only even powers of $q$.
Over the measured wavevector range, we approximate $\widetilde{c}_P(q)$ by an even polynomial,
\begin{equation}
\widetilde c_P(q) = c_0 + c_1q^2 + c_2q^4 + c_3q^6, \label{eq:cP}
\end{equation}
where coefficients $c_0$, $c_1$, $c_2$, and $c_3$ are fitted to passive simulations separately for each pair $(\rho_0,D_t)$.
Importantly, $c_0$, $c_1$, $c_2$, and $c_3$ are kept fixed when activity is switched on ($\text{Pe}>0$). 

Consequently, the active structure factor $S(q)$ follows from Eq.~(\ref{eq:Sq}) under the substitution~(\ref{eq:passive-replacement}), and the active intermediate scattering function $F(q,t)$ follows from Eqs.~(\ref{eq:Fq-1})--(\ref{eq:Fq-3}) using both the modified $S(q)$ and the modified eigenvalues $\lambda_\pm$.

Fig.~\ref{fig:WCA-static} compares the fitted analytic structure factor $S(q)$ (solid black lines) with the simulations (markers) for different three different values of $D_t$: (a) $D_t=0.05$, (b) $D_t=0.5$, and (c) $D_t=5$, while $\rho_0=0.2$ is fixed.
The four coefficients at each value of $D_t$ are fitted using passive data at $\mathrm{Pe}=0$ and retained to obtain the active curves at $\mathrm{Pe}=0.7$, $1.3$, and $2.0$ (solid black lines). 
This procedure uses less structural information than orientation-resolved active pair-distribution closures~\cite{JeggleStenhammarWittkowski2020,BrokerEtAl2024}.
Using static correlations as input to a dynamic theory is also central to MCT~\cite{Janssen2018,Liluashvili2017,Szamel2019,DebetsEtAl2024}. 

Finally, using the same fitted parameters $c_0$, $c_1$, $c_2$, and $c_3$, we obtain the prediction for the intermediate scattering function $F(q,t)$ for both passive $\text{Pe}=0$ and active $\text{Pe}>0$.
Using the same four coefficients, we can also predict the critical $\text{Pe}_c$ for the onset of oscillations.
Fig.~\ref{fig:WCA-intermediate} shows the comparison between these predicted curves (solid black lines) and the simulation data (markers) for three different values of wavectors: (a) $q=0.0320$, (b) $q=0.0915$, and (c) $q=0.149$, with fixed $\rho_0=0.2$ and $D_t=0.05$.
The corresponding critical $\text{Pe}_c$ are $\text{Pe}_c\simeq22$ for (a), $\text{Pe}_c\simeq7.7$ for (b), and $\text{Pe}_c\simeq4.7$ for (c).
Our prediction qualitatively captures the crossover to the oscillatory relaxation at high activity $\text{Pe}>\text{Pe}_c$.

%%%
\section{Conclusions}\label{sec:conclusions}
%%%

Using the Dean--Kawasaki framework with an angular-moment closure, we have derived fluctuating hydrodynamic equations for the density and polarization fields. 
Within the Gaussian theory approximation about a homogeneous isotropic state, we obtain analytical expressions for the static structure factor $S(q)$ and the collective intermediate scattering function $F(q,t)$. 
For Gaussian-core interactions and any other well-behaved potentials with $V'(r\rightarrow0)=0$, the analytic predictions depend only on the microscopic parameters and reproduce both $S(q)$ and $F(q,t)$ at low activity and long wavelengths without any fitting parameters.

Retaining the polarization dynamics also allows the theory to predict a crossover from exponential relaxation to damped oscillations above the wavevector-dependent activity threshold $\mathrm{Pe}_c(q)$ given in Eq.~(\ref{eq:Pe-critical}). 
These oscillations arise from persistent self-propulsion and the coupling between density and longitudinal polarization, despite the absence of inertia. 
The predicted crossover is reproduced by the simulations.

For the singular WCA potential, we replace the interaction potential $-\widetilde{V}(q)/D_T$ with the passive direct-correlation function, which is approximated by an even sixth order polynomial that is fitted to the passive data.
The polynomial representation has four coefficients, which we fit to passive simulation data separately for each pair of density and diffusivity, then held fixed as activity is varied.
The same passive coefficients determine both $S(q)$ and $F(q,t)$ without additional activity-dependent fitting parameters. 
The resulting theory qualitatively captures the crossover to oscillatory relaxation.

The Gaussian theory further yields the bare response propagator.
Together with the Gaussian noise covariance, it provides a starting point for a diagrammatic perturbation theory in which the neglected nonlinear deterministic and multiplicative-noise terms generate interaction vertices.
A loop expansion could then be used to calculate corrections to the response and density correlations.
The importance of higher-order interactions is illustrated in~\cite{GarciaMillanEtAl2026}, 
which demonstrates how effective attraction may emerge beyond the leading-order repulsive contribution for two soft run-and-tumble particles.
Extending the present ABP theory in this direction, and retaining additional angular moments where necessary, could help resolve the discrepancies observed at larger activity and possibly capture MIPS.

\begin{acknowledgments}
SC and ET acknowledges funding from EPSRC grant no.~EP/W027194/1. 
AS and TM acknowledge funding from the Israel Science Foundation (Grant No.~1356/22)
\end{acknowledgments}

\section*{Author declarations}
\subsection*{Conflict of interest}
The authors have no conflicts to disclose.
% AUTHOR CHECK: add an agreed CRediT author-contributions statement. Contributions cannot be inferred from the supplied files.

\section*{Declaration of generative AI use}
During the preparation of this manuscript, the authors used ChatGPT (OpenAI) to assist with literature searches and improve the language and clarity of the text. 
The authors verified the references, reviewed and edited the resulting text, and take full responsibility for the content of the manuscript.

\section*{Data availability}
The data that support the findings of this study are available from the corresponding author upon request.

%%%%%
\appendix
%%%%%

%%%
\section{Nondimensionalization}\label{sec:nondimensionalization}
%%%

In this section we use overbars for dimensional quantities. Quantities without overbars are dimensionless.
The overdamped equations for $N$ Active Brownian Particles (ABPs) are
\begin{align}
  \frac{d\bar{\bm{r}}_i}{d\bar{t}} & = \bar{v}_0 \hat{\boldsymbol{\nu}}_i 
  			- \frac{1}{\bar{\gamma}} \sum_{j\neq i} \bar{\boldsymbol\nabla}_i \bar{V}(|\bar{\mathbf{r}}_i - \bar{\mathbf{r}}_j|)
			+ \sqrt{2\bar{D}_t}\bar{\bm{\xi}}_i(\bar{t}), \label{eq:rdot-dim} \\
  \frac{d\theta_i}{d\bar{t}} &= \sqrt{2 \bar{D}_r}\bar{\zeta}_i(\bar{t}),  \label{eq:thetadot-dim}
\end{align}
where $\bar{\gamma}$ is the translational friction coefficient, $\bar{D}_t$ is the translational diffusion constant, and $\bar{D}_r$ is the rotational diffusion constant.
If Stokes-Einstein relation holds, we also have $\bar{D}_t=\bar{k}_B\bar{T}/\bar{\gamma}$, where $\bar{T}$ is the temperature and $\bar{k}_B$ is the Boltzmann constant.
Using the length scale $\bar{\sigma}$ (e.g. diameter of particle), energy scale $\bar{\epsilon}$ (e.g. typical interaction energy), and time scale $\bar{t}_0=\bar{\gamma}\bar{\sigma}^2/\bar{\epsilon}$, 
we define the dimensionless position, potential, and time as
\begin{equation}
\mathbf{r}_i = \frac{\bar{\mathbf{r}}_i}{\bar{\sigma}}, \quad 
V(r) = \frac{\bar{V}(\bar{\sigma}r)}{\bar{\epsilon}}, \quad\text{and}\quad 
t = \frac{\bar{t}}{\bar{t}_0}, \label{eq:dimensionless}
\end{equation}
respectively.
The dimensionless self-propulsion speed, translational diffusion, and rotational diffusion are given by
\begin{equation}
v_0 = \frac{\bar{v}_0\bar{t}_0}{\bar{\sigma}}, \quad 
D_t = \frac{\bar{D}_t\bar{\gamma}}{\bar{\epsilon}}, \quad\text{and}\quad 
D_r = \bar{D}_r\bar{t}_0,
\end{equation}
respectively.
The dimensionless noise are given by
\begin{equation}
\boldsymbol{\xi}_i(t) = \sqrt{\bar{t}_0}\bar{\boldsymbol{\xi}}_i(\bar{t}_0 t) \quad\text{and}\quad
\zeta_i(t) = \sqrt{\bar{t}_0}\bar{\zeta}_i(\bar{t}_0 t).
\end{equation}
Additionally, we also define the dimensionless average number density as $\rho_0=N/L^2$, where $L=\bar{L}/\bar{\sigma}$ is the dimensionless system size.

%%%
\section{Derivation of Dean--Kawasaki equation}\label{sec:DK}
%%%

In this section we derive the distributional Dean--Kawasaki equation in (\ref{eq:psidot}).
Let us consider a general function of the position and orientation of a tagged particle $i$, denoted by $f(\mathbf{r}_{i}(t),\theta_{i}(t))$.
The rate of change of $f$ with respect to time $t$ is given by the It\^{o}'s lemma for the chain rule
\begin{align}
\frac{df}{dt} = & \Bigg[ v_{0}\hat{\boldsymbol{\nu}}_{i} - \sum_{j\neq i} \boldsymbol{\nabla}_i V(|\mathbf{r}_{i}-\mathbf{r}_{j}|) 
			+ \sqrt{2D_{t}}\boldsymbol{\xi}_{i} \Bigg] \cdot \boldsymbol{\nabla}_i f \nonumber \\
	&+ D_{t}\nabla^{2}_{i}f+\sqrt{2D_{r}}\zeta_{i} \frac{\partial f}{\partial\theta_{i}}
	+ D_{r}\frac{\partial^{2}f}{\partial\theta^{2}_{i}}. \label{eq:Ito-lemma}
\end{align}
Note that repeated particle indices $i$ are not summed. 
Next, we introduce the distribution of the tagged particle $i$ as 
\begin{equation}
\psi_{i}(\mathbf{r},\theta,t) = \delta(\mathbf{r}-\mathbf{r}_{i}(t))\delta(\theta-\theta_{i}(t)).
\end{equation}
Eq.~(\ref{eq:Ito-lemma}) can then be written as
\begin{align}
\frac{df}{dt} &= \int d\mathbf{r}\int_0^{2\pi} d\theta \nonumber\\
\Bigg\{ & \boldsymbol{\nabla} \cdot 
		\Bigg[ -\psi_iv_{0}\hat{\boldsymbol{\nu}} + \psi_i\sum_{j\neq i} \boldsymbol{\nabla} V(|\mathbf{r} - \mathbf{r}_{j}|) - \sqrt{2D_{t}}\boldsymbol{\xi}_i\psi_i \Bigg] \nonumber\\
 &+ D_t\nabla^2 \psi_i - \sqrt{2D_r}\zeta_i \partial_\theta\psi_i + D_r\partial^2_\theta\psi_{i} 
 	\Bigg\} f, \label{eq:fdot-ABP-1}
\end{align}
where we have denoted $\hat{\boldsymbol{\nu}}(\theta)=(\cos\theta,\sin\theta)^{T}$ and used integration by parts. 
We also know that
\begin{equation}
\frac{df}{dt}=\int d\mathbf{r}\int_0^{2\pi} d\theta\,f(\mathbf{r},\theta)\partial_t\psi_{i}(\mathbf{r},\theta,t). \label{eq:fdot-ABP-2}
\end{equation}
Comparing Eq.~(\ref{eq:fdot-ABP-1}) to Eq.~(\ref{eq:fdot-ABP-2}), and rewriting
\begin{equation}
\boldsymbol{\nabla}V(|\mathbf{r} - \mathbf{r}_{j}|) = \int d\mathbf{r}'\int_0^{2\pi} d\theta' \, \psi_{j}(\mathbf{r}',\theta',t)\boldsymbol{\nabla}V(|\mathbf{r}-\mathbf{r}'|),
\end{equation}
we obtain
\begin{align}
 \partial_t\psi_i 
=& -\boldsymbol{\nabla} \cdot \Bigg[ 
	\psi_i v_0 \hat{\boldsymbol{\nu}} + \sqrt{2D_{t}}\boldsymbol{\xi}_{i}\psi_{i} \nonumber \\
  &\quad -\psi_i\sum_{j\neq i} \int d\mathbf{r}'\int_0^{2\pi} d\theta' \, \psi_{j}(\mathbf{r}',\theta',t)\boldsymbol{\nabla}V(|\mathbf{r}-\mathbf{r}'|)
	 \Bigg] \nonumber \\
 & +D_{t}\nabla^{2}\psi_{i} - \sqrt{2D_{r}}\zeta_{i}\partial_\theta\psi_i + D_{r}\partial^2_\theta\psi_i. \label{eq:psidot_i}
\end{align}
Now we define the empirical position-orientation distribution function as [cf. Eq.~(\ref{eq:psi})]:
\begin{equation}
\psi(\mathbf{r},t) = \sum_{i=1}^N \psi_i = \sum_{i=1}^N \delta(\mathbf{r}-\mathbf{r}_i(t))\delta(\theta-\theta_i(t)).
\end{equation}
Summing Eq.~(\ref{eq:psidot_i}) over $i$ yields the time-evolution equation for $\psi$.
The interaction term contains the restricted sum $\sum_{i}\sum_{j\neq i}$, which we rewrite as the unrestricted double sum $\sum_{i}\sum_{j}$ minus the diagonal contribution with $i=j$.
The diagonal contribution $i=j$ gives rise to the self-interaction correction term $\mathbf{S}_0$ in Eq.~(\ref{eq:psidot}) of the main text.

%%%
\section{Two-component complex Ornstein--Uhlenbeck process}\label{sec:2d-OU}
%%%

In this section we outline the solution to the two-component complex Ornstein--Uhlenbeck process, such as the one in Eq.~(\ref{eq:Gaussian-theory}).
For clarity of presentation, we drop the subscript $\mathbf{q}$ in this section.
A general two-component complex Ornstein-Uhlenbeck process is described by the following equation:
\begin{equation}
\frac{dX_{a}(t)}{dt}=-M_{ab}X_{b}(t)+Z_{a}(t),\label{eq:Xdot-1}
\end{equation}
where $a,b\in\{1,2\}$. 
$\mathbf{Z}(t)$ is a Gaussian white noise with zero mean and Dirac delta correlation
\begin{equation}
\left\langle \mathbf{Z}(t)\mathbf{Z}^{\dag}(t')\right\rangle_{ab} = \left\langle Z_{a}(t)Z_{b}(t')^*\right\rangle = L_{ab}\delta(t-t').
\end{equation}
The symbol $^\dag$ denotes the conjugate transpose of a matrix.
Suppose we discretize the time into $t=n\Delta t$, where $n\in\{0,1,2,\dots\}$ and $\Delta t$ is the time step. 
We can then write Eq.~(\ref{eq:Xdot-1}) to lowest order in discretized time as~\cite{SorkinArielMarkovich2025}
\begin{align}
X_{a}(t+\Delta t) = X_{a}(t) -\Delta tM_{ab}X_{b}(t)+\sqrt{\Delta t}\bar{Z}_{a}(t), \label{eq:delta-X}
\end{align}
where $\bar{Z}_{a}(t)$ is the discretized noise with Kronecker delta correlation
\begin{equation}
\left\langle \bar{Z}_{a}(t)\bar{Z}_{b}(t')^*\right\rangle = L_{ab}\delta_{t,t'}.
\end{equation}
The symbol $^*$ indicates complex conjugation.

We define the stationary equal time covariance to be:
\begin{equation}
C_{ab}(0)=\left\langle X_{a}(\tau)X_{b}(\tau)^*\right\rangle _{\text{ss}},
\end{equation}
where the average is taken in the stationary state, i.e. $\tau\rightarrow\infty$.
Throughout this paper, we have assumed the existence of a stationary state.
To find $C_{ab}(0)$,
we first consider the rate of change of $\left\langle X_{a}(\tau)X_{b}(\tau)^*\right\rangle$ with respect to $\tau$.
First, let us look at the variation
\begin{align}
& \left\langle X_{a}(\tau+\Delta t)X_{b}(\tau+\Delta t)^*\right\rangle - \left\langle X_{a}(\tau)X_{b}(\tau)^*\right\rangle   \nonumber\\
 &= -\Delta t\left\langle X_{a}(\tau)X_{c}(\tau)^*\right\rangle M^{\dag}_{cb} \nonumber\\
 &\quad - \Delta tM_{ac}\left\langle X_{c}(\tau)X_{b}(\tau)^*\right\rangle \nonumber\\
 &\quad + \Delta t\underbrace{\left\langle \bar{Z}_{a}(\tau)\bar{Z}_{b}(\tau)^*\right\rangle }_{L_{ab}} + \mathcal{O}(\Delta t^{3/2}). \label{eq:variation-C0}
\end{align}
Note that in this derivation, we have additive noise, for which $\mathbf{L}$ does not depend on $\mathbf{X}$.
In the case of multiplicative noise $\mathbf{L}(\mathbf{x})$, the above expression is still valid provided the It\^{o} interpretation is used~\cite{SorkinArielMarkovich2025}.
Dividing Eq.~(\ref{eq:variation-C0}) by $\Delta t$ and taking the limit $\Delta t\rightarrow0$, we obtain
\begin{align}
& \frac{d\left\langle X_{a}(\tau)X_{b}(\tau)^*\right\rangle }{d\tau} \nonumber\\
&= -\left\langle X_{a}(\tau)X_{c}(\tau)^*\right\rangle M^{\dag}_{cb}-M_{ac}\left\langle X_{c}(\tau)X_{b}(\tau)^*\right\rangle + L_{ab}.
\end{align}
In the stationary state $\tau\rightarrow\infty$, the left hand side of the above equation is zero, and we have
\begin{equation}
\mathbf{C}(0)\mathbf{M}^{\dag} + \mathbf{M}\mathbf{C}(0) = \mathbf{L}.
\end{equation}
Solving these simultaneous equations determines each element of the matrix $\mathbf{C}(0)$, although only $C_{11}(0)$ is needed to obtain the static structure factor $S(q)$ in the main text.

Now we define the unequal time covariance as:
\begin{equation}
C_{ab}(t) = \left\langle X_{a}(\tau+t)X_{b}(\tau)^*\right\rangle _{\text{ss}},
\end{equation}
where the angle bracket is a stationary average and we consider the limit $\tau\rightarrow\infty$. 
The rate of change of $C_{ab}(t)$ with respect to $t$ is given by (for $t\ge0$)
\begin{align}
\frac{dC_{ab}(t)}{dt} =& \left\langle \left.\frac{dX_{a}}{dt}\right|_{\tau+t}X_{b}(\tau)^*\right\rangle _{\text{ss}} \\
	= &- \left\langle M_{ac}X_{c}(\tau+t)X_{b}(\tau)^*\right\rangle _{\text{ss}} \nonumber\\
	   &\quad + \left\langle Z_{a}(\tau+t)X_{b}(\tau)^*\right\rangle \\
	 =& -M_{ac}C_{cb}(t), \label{eq:Cdot}
\end{align}
where we have used the fact that our stochastic equation (\ref{eq:delta-X}) is written in the It\^{o} convention such that the noise $Z_a(\tau+t)$ is independent of $X_b(\tau)$ for $t\ge0$.
The formal solution to Eq.~(\ref{eq:Cdot}) is 
\begin{equation}
\mathbf{C}(t)=e^{-\mathbf{M}t}\mathbf{C}(0),\quad\text{for }t\ge0.
\end{equation}
To evaluate the matrix exponential, we seek eigenmode solutions of the form
\begin{equation}
\begin{pmatrix}
C_{1b}(t)  \\
C_{2b}(t) 
\end{pmatrix}
\propto \mathbf{v} e^{-\lambda t},
\end{equation}
where $b\in\{1,2\}$ and $\mathbf{v}$ is an eigenvector of $\mathbf{M}$ with eigenvalue $\lambda$, satisfying $\mathbf{M}\mathbf{v}=\lambda\mathbf{v}$.
When $\mathbf{M}$ has two linearly independent eigenvectors $\mathbf{v}_\pm$ with eigenvalues $\lambda_\pm$, the solution is the linear combination of the two eigenmodes: 
\begin{equation}
\begin{pmatrix}
C_{1b}(t)  \\
C_{2b}(t) 
\end{pmatrix}
= A \mathbf{v}_+ e^{-\lambda_+ t} + B \mathbf{v}_- e^{-\lambda_- t} ,
\end{equation}
where $A$ and $B$ are constants of integration, set by the initial condition $\mathbf{C}(0)$.

%%%
\section{Numerical computation of $S(q)$ and $F(q,t)$}\label{sec:Sq-Fq}
%%%

The intermediate scattering function $F(q,t)$ can be expressed in terms of the particles' positions $\{\mathbf{r}_i(t)\}$.
We define the density fluctuation as
\begin{align}
\delta\rho(\mathbf{r},t) = \sum_{i=1}^N \delta(\mathbf{r}-\mathbf{r}_i(t)) - \rho_0, \label{eq:deltarho-particle-1}
\end{align}
Using the definition of the Fourier transform in (\ref{eq:deltarho-FT}), we get
\begin{equation}
\delta\tilde{\rho}_{\mathbf{q}}(t) = \frac{1}{L}\sum_{i=1}^N e^{-i\mathbf{q}\cdot\mathbf{r}_i(t)} - \frac{N}{L}\delta_{\mathbf{q},0}. \label{eq:deltarho-particle}
\end{equation}
Substituting (\ref{eq:deltarho-particle}) into the definition of the intermediate scattering function (\ref{eq:Fq-def}), we obtain
\begin{align}
&F(q,t) \nonumber\\
&=  \frac{1}{N} \Bigg\langle 
		\left( \sum_{i}\cos\left[\mathbf{q}\cdot\mathbf{r}_{i}(\tau+t)\right] \right)
		\left( \sum_{j}\cos\left[\mathbf{q}\cdot\mathbf{r}_{j}(\tau)\right] \right) \nonumber\\
	  &\quad + \left( \sum_{i}\sin\left[\mathbf{q}\cdot\mathbf{r}_{i}(\tau+t)\right] \right)
	     \left( \sum_{j}\sin\left[\mathbf{q}\cdot\mathbf{r}_{j}(\tau)\right]\right)
	     \Bigg\rangle_{\text{ss}} \nonumber\\
    &\quad-N\delta_{\mathbf{q},0}.
\end{align}
To calculate $S(q)$, we set $t=0$ in the expression above. 
In numerical simulations, $\left<\dots\right>_{\text{ss}}$ indicates time and/or ensemble averaging in the steady state.
For the time average, we average over different reference times $\tau$'s, that are large enough compared to the relaxation timescale of the system.

\bibliographystyle{aipnum4-2}
\bibliography{bib}
\end{document}